\documentclass[11pt]{article}
\usepackage{geometry}                
\usepackage{url}  
\usepackage{graphicx}
\usepackage{amssymb}
\usepackage{epstopdf}
\usepackage{amsmath}
\usepackage{amscd}
\usepackage{amsthm}
\usepackage{mathrsfs}
\usepackage{amscd}
\usepackage{color}
\usepackage[T1]{fontenc}
\usepackage[english]{babel}
\usepackage{tikz,tikz-3dplot}
\usepackage{bm}
\usepackage{xcolor}
\usepackage[normalem]{ulem}
\usepackage{algorithm}
\usepackage{algpseudocode}
\usepackage{subcaption}
\usepackage{wrapfig}
\usepackage{lscape}
\usepackage{longtable}
\usepackage{setspace}
\usepackage{xspace}
\usepackage[round]{natbib}
\usepackage{arydshln}

\newcommand{\snapper}{S\textsc{napper}\xspace}
\newcommand{\snapp}{S\textsc{napp}\xspace}
\newcommand{\sX}{\mathcal{X}}

\newcommand{\sfrac}[2]{\mbox{$\frac{#1}{#2}$}}
\newcommand{\half}{\sfrac{1}{2}}

\title{Bayesian inference of species trees using diffusion models}
\author{Marnus Stoltz$^1$ \and Boris Bauemer$^1$ \and Remco Bouckaert$^2$ \and Colin Fox$^3$ \hspace{0.8cm} Gordon Hiscott$^1$ \and David Bryant$^{1*}$\\
{\small $1.$ Department of Mathematics and Statistics, University of Otago}\\
{\small $2.$ Centre for Computational Evolution, University of Auckland}\\
{\small $3.$ Department of Physics, University of Otago.}\\
{\small $*$ Corresponding author.   {\tt david.bryant@otago.ac.nz}}
}

\date{\today}

\begin{document}

\maketitle

\begin{abstract}
We describe a new and computationally efficient Bayesian methodology for  inferring species trees and demographics from unlinked binary markers. Likelihood calculations are carried out using diffusion models of allele frequency dynamics combined with a new algorithm for numerically computing likelihoods of quantitative traits. The diffusion approach allows for analysis of datasets containing hundreds or thousands of individuals. The method, which we call \snapper,  has been implemented as part of the Beast2 package. We introduce the models, the efficient algorithms, and report performance of \snapper on simulated data sets and on SNP data from rattlesnakes and freshwater turtles. 
\end{abstract}

\section{Introduction}
Recent years have witnessed a proliferation in the number of methods for inferring species trees from whole genomes \citep{BryantETAL2019a}. 
Some have gone so far as to describe this as a {\em paradigm} shift in phylogenetics \citep{edwards2009new}. It is now widely accepted that (i) phylogenetic analysis needs to take account of the varying evolutionary histories of different parts of the genome, and (ii) estimation of evolutionary relationships between populations (or species) should take account of evolutionary dynamics  {\em within} populations (or species).

In the systematics community, taking account of population dynamics in a phylogenetic context has meant implementing some version of the {\em multispecies coalescent} \citep{liu2008estimating, liu2008best,larget2010bucky, heled2013simulating}. The coalescent models the distribution of gene trees within a population; the multispecies coalescent is its natural extension to multiple populations or species. One of the key advantages of the coalescent is that, for neutral mutations, the gene tree can be decoupled from the mutation process, a feature which forms the basis of many implementations of the model.
Nevertheless, the coalescent has its limitations. It is difficult to incorporate selection, for example. Also, the running time of coalescent based methods (e.g. BEAST, *BEAST, etc.) depends critically on the number of individuals being sampled. In *BEAST a gene tree is sampled for each loci, with the number of leaves in the gene tree given by the number of individuals sampled from all populations. 

\citet{Bryant2012} showed that explicit sampling of gene trees for each locus could be avoided in the case of unlinked binary markers. This is appropriate for the estimation of species trees from unlinked single nucleotide polymorphisms (SNPs). Their algorithm was implemented in \snapp, and can analyse data sets with hundreds of thousands of loci and up to 200  individuals (depending on computing resources and sampling difficulties). Unfortunately \snapp does not scale that well as the number of individuals increases. The running time of the algorithm algorithm is $O(n^2 \log n)$ with $n$ individuals, which quickly becomes limiting. The far most common user complaint about \snapp to developers has been running speed. 
 
Rather than tinker with the \snapp algorithm, we have taken a completely new approach with a different kind of model. Like \citet{gutenkunst2010diffusion}, \citet{siren2010reconstructing} and \citet{Hey2012} we use diffusion models, though we apply them in a new way. The net result is that we can carry out \snapp-type analyses but with essentially no limits on the number of individuals being sampled.

Diffusion models, like the coalescent, are a convenient approximation of the standard Wright-Fisher discrete models. In fact, diffusion models pre-date coalescent models by a good fifty years \citep{wright1931evolution,Kingman1982b}. Whereas coalescent models describe the distribution of the genealogy or gene tree for a sample of individuals, diffusion models describe the frequency of an allele in the population as a whole. There are straightforward extensions incorporating selection.

There are three challenges to overcome working with a model of allele frequencies. First, we do not actually observe the population frequencies, we just observe a sample drawn from that population. This problem is easily solved by incorporating the sampling step explicitly into our likelihood.

Second, gene frequencies are continuous traits, so it is not possible to sum over ancestral trait values as in Felsenstein's pruning algorithm \citep{Felsenstein1981b}. For this, we follow a numerical approach developed in \citet{Hiscott2016}, though with some new twists to improve efficiency and accuracy. 

Third, diffusion models do not give explicit transition probabilities or densities: these are only available via partial differential equations (PDEs). We solve these differential equations numerically, extending a standard spectral approach from a single population to the entire species tree. We note that there is significant potential for extending our methods to other diffusion models.

The outline of this paper is as follows: In Section 2 we briefly review the population genetic diffusion models and some relevant theory, describe the model on a species tree and discuss other related models. In Section 3 we spend some time setting up the mathematics developed to compute the model likelihood. Thereafter we present a numerical algorithm for computing the likelihood and discuss the Bayesian software package \snapper.  In Section 4 we assess the perfomance of \snapper and compare running times to \snapp. In Section 5 we showcase the capability of \snapper on a SNP dataset from freshwater turtles. We conclude the paper with a summary of method implementation and discuss some possible model extensions for future work. 
\section{Models and Methods}

\subsection{The Wright-Fisher diffusion}

Our starting point is the Wright-Fisher discrete model of drift and mutation. Suppose we have a population of $N$ diploid individuals, giving $2N$ copies of each autosomal gene. We consider a gene with two alleles (say `red' and `green') and, to being with, a model with non-overlapping generations. With random mating the number $X_{n+1}$ of red alleles in generation $n+1$ has a binomial distribution with parameters $2N$ and $(1-u) \frac{X_n}{2N} + v \left(1- \frac{X_n}{2N}\right)$, where $u$ is the probability of mutating from red to green and $v$ the probability of mutating from green to red. In this way the number of red alleles will follow a random walk with discrete time (generation number) and states (from $0$ to $2N$). 

The idea behind the diffusion models is to approximate this discrete random walk by a continuous random walk that is easier to work with analytically. We set up the approximation so that the larger $N$ gets, the better the approximation fits. We describe the {\em proportion} of gene copies having the red allele, rather than the total count.  The state space of the process will be the interval $[0,1]$. Instead of considering discrete generations, we construct a random walk which is continuous in time.

Diffusion models for gene frequencies involve a rescaling of time. There are simple, practical, reasons for this. As the population size $N$ gets larger, the rate of genetic drift decreases, ultimately approaching zero drift in the limit. However the effect of drift is something that we would like to model. For this reason we change the units of time so that the rate of drift remains approximately the same as $N$ increases, eventually converging to some non-zero amount. The convention for diploid populations is to use a scale where $1$ unit of time corresponds to $2N$ generations (one coalescent unit). 

If we are to change the units of time, we need to adjust the rate of mutation accordingly. Therefore the overall rate of change due to mutations from green alleles to red alleles is $2Nu$. We adopt the standard notation and define $\beta_1 = 2Nu$ and $\beta_2 = 2Nv$. 

For each value of $t$ we let $f(x,t)$ denote the {\em density} of the allele proportion at time $t$, noting that this allele proportion is a continuous random variable with state space $[0,1]$. Surprisingly, there is very little choice over what the function $f(x,t)$ might be, after a few basic assumptions are made. See \citet{Etheridge2011} for technical details and \citet{McKane2007} for a discussion about how we need to incorporate point masses at $0$ and $1$ into $f(x,t)$.

As is often the case in mathematical modelling, we work with the function $f$ indirectly using a PDE.
Stochastic process theory \citep[chap. 5]{oksendal2003stochastic}    tells us that the function $f$ satisfies the PDE
\begin{align}
\frac{df(x,t)}{dt} &=  -\frac{d}{dx} \left( \beta_1(1-x) -\beta_2x \right) f(x,t)+ \frac{1}{2}  \frac{d^2}{dx^2}x(1-x)f(x,t). \label{eq:forward}
\end{align}
This equation by itself is not enough to uniquely determine $f$. We also need to specify what $f$ looks like at the boundaries. To specify that the distribution of the initial state is given by some density $\pi$ we add the initial condition
\begin{equation}
f(x,0)  = \pi(x) \mbox{ for $x \in [0,1]$ }.\label{eq:initialForward}
\end{equation}
Note that to specify the initial value exactly, we need $\pi$ to be a Dirac-delta function $\pi(x) = \delta(x - x_0)$ which is essentially a infinitely thin spike on one value $x_0$. Even with these initial conditions, the PDE \eqref{eq:forward} does not uniquely determine the function $f$. We also need to add boundary conditions at $x=0$ and $x=1$ to guarantee that the probability of going outside the interval $[0,1]$ is always zero. Writing this as a condition on integrals of $f$ and then plugging  in the PDE, eventually leads to 
what in physics is known as a {\em zero-flux condition} 
\begin{equation}
-(\beta_1(1-x) -\beta_2x)f(t,x) + \frac{1}{2} \frac{d}{dx}x(1-x)f(t,x)  = 0, \label{eq:noFlux}
\end{equation}
when $x = 0$ or $x=1$. See \citet{McKane2007} for details.

The PDE \eqref{eq:forward} together with the boundary conditions \eqref{eq:initialForward} and \eqref{eq:noFlux} are just a mathematically convenient, if not particularly transparent, way to describe the function $f$. The function $f$, in turn, is just a way to approximate the probabilities in the original discrete process. The diffusion approximation works well only if mutations rates are assumed small to begin with. This will give a reasonably small error in the diffusion approximation. As shown in \citet{ethier1977error} we still have convergence in expectation with an error bound on the diffusion approximation behaving like $O(u+v+1/N)$. Note that if $u$ or $v$ are large (compared to, say, $\frac{1}{N}$) then the diffusion approximation will fail miserably. In a recent simulation study, \citet{tataru2017statistical} used simulations to quantify the error from the diffusion approximation in small populations, and found that diffusion models gave reasonable approximations even when the population has fewer than 100 individuals. 

\subsection{Modelling allele frequencies on a species tree}

The diffusion model describes how allele frequencies change over time in a single population. The model extends directly to multiple populations in 
a species tree \citep{siren2010reconstructing}. As in \citet{Bryant2012} we think of the root of the species tree to be at the top of the tree and the leaves at the bottom. The model describes evolution of the allele frequencies from the top of the tree to the bottom. 

The allele frequency at the root has a distribution given by the stationary density of the diffusion model (in this case, a beta distribution). The allele frequency at the bottom of a branch has a distribution given by the diffusion model with an initial value equal to the allele frequency at the top of the branch. At a speciation, the two daughter populations have the same allele frequency as the parent population. 

We now formalise these ideas.  Suppose that the branches in the species tree are numbered $i=1,2,\ldots,2n-2$ where, for convenience, the branches adjacent to the leaves are numbered $1,2,\ldots,n$.  Let  $\sX^T_i$ denote the allele frequency immediately below the top of branch $i$. Let $X^B_i$ denote the allele frequency immediately above the bottom of branch $i$. Let $X_\rho = X_\rho^B$ denote the allele frequency at the root. 

At the root, $\rho$, the proportion of red alleles in the ancestral population equals the stationary distribution of the diffusion model. The stationary distribution of the diffusion equation \eqref{eq:forward} has a beta distribution \citep{Ewens2004} with density
\begin{equation}
\pi_\rho(x|\beta_1, \beta_2) = \frac{\Gamma\{2\beta_1 + 2 \beta_2\}}{\Gamma\{2 \beta_1\}\Gamma\{2 \beta_2\}} x^{2\beta_1 - 1}(1-x)^{2\beta_2-1}, \qquad 0<x<1
\label{eqn: stationary distr.}
\end{equation}
Along any branch in the species tree, the changes in allele frequencies are modelled using the diffusion process. The allele frequency at the start of the branch gives the initial density, $f(x,0)$, for $x\in [0,1]$. The distribution of the allele frequency $y$ at the end of the branch is then given by $f(y,t)$, where $t$ is the length of the  branch in units of $2N$ generations and $y \in [0,1]$. 

At a speciation, we make the assumption that there is no correlation between allele type and speciation. Hence the two descendant species are assumed to have identical allele frequencies to the parent node.

We have, therefore, a model for allele frequencies over the whole tree. However allele frequencies at the leaves are not directly observed. Instead we have a sample of $n_i$ individuals, giving $2n_i$ allele copies taken from each species $i$. If $x_i$ is the red allele frequency for the population then the observed number $r_i$ of red alleles in the the sample for this gene has binomial distribution with parameters $2n_i$ and $x_i$, so that
\[P[R_i=r | x_i] = {2n_i \choose r} x_i^r (1-x_i)^{2n_i-r},  \]       
see \citet{siren2010reconstructing}.

In Algorithm \ref{alg:simulate_allele_freq} we describe how to simulate allele frequencies under this model. Note that in a {\em pre-order traversal} we visit every node in order so that the children of a node are always visited {\em after} the node itself.

\begin{algorithm}
	\caption{Simulate allele frequencies on a tree}
	\label{alg:simulate_allele_freq}
	\begin{algorithmic}[1]
		\Procedure{TreeSim}{S}
		\State Generate $X^B_{\rho}$ from density $\pi(x|\beta_1,\beta_2)$ given in (\ref{eqn: stationary distr.})
		\For{all nodes $i$ except the root in a pre-order traversal}
		\State $X_{i}^T \leftarrow X_{p(i)}^B$, where $p(i)$ denotes the parent of $i$
		\State Use diffusion to simulate $X^B_i$ given initial value $X^T_i$  
		\If{node $i$ is a leaf}
		\State Generate $r_i$  from $Bin(2n_i,X_{i}^B)$   
		\EndIf		
		\EndFor
		\EndProcedure
	\end{algorithmic}
\end{algorithm}

We use the algorithm of \citet{jenkins2017exact} to simulate diffusions along each branch. Alternative methods for simulating diffusions include truncating  a  spectral  expansion of the transition function \citep{Song2012,Steinrucken2014} or numerical solutions of the Kolmogorov equations \citep{williamson2005simultaneous,bollback2008estimation,Gutenkunst2009}. Care needs to be taken when implementing these methods since approximation errors close to the boundary can become large, a problem that \citet{jenkins2017exact}  explicitly avoid.

\subsection{Analytical formula for partial likelihoods}

We now describe how to describe the likelihood functions analytically. Overall our approach is to use dynamic programming, as in Felsenstein's pruning algorithm \citep{Felsenstein1981b}. For each node $i$ and each ancestral state $x$, which is our case is an allele frequency, we define two partial likelihoods, $\ell^B_i(x)$ and $\ell^T_i(x)$. The first is the probability of observing all allele counts for taxa in the subtree rooted at $i$, conditional on the state $X^B_i$ at the bottom of branch $i$ being equal to $x$. The second is defined in the same way, but is instead conditioned on the state $X^T_i$ at the top of the branch.  A similar approach was used in \citet{Bryant2012}.

We now show how to describe these partial likelihoods recursively for an individual site $m$.
 
\subsubsection{Partial likelihoods at the leaves}

Suppose that node $i$ is a leaf and that we have sampled $n_i$ diploid individuals from this population. Let $r_i$ denote the number of observed gene copies carrying the red allele for site $m$. If $x$ is the proportion of red alleles in the population then $r_i$ has a binomial distribution with parameters $n_i$ and $x$. Hence the partial likelihood at a leaf $i$ is given by
\begin{equation}
\ell_i^B(x) = {2n_i \choose r_i} x^{r_i}(1-x)^{2n_i - r_i}.
\label{eqn: like-tip}
\end{equation} 

\subsubsection{Partial likelihoods at a speciation}

Let $j$ and $k$ be the children of node $i$. We assumed that the allele frequencies for daughter populations after a speciation are exactly those of the parent population before speciation. The partial likelihoods at the bottom of the branch above $i$ is then the product of partial likelihoods at the tops of the branches immediately below $i$, so
\begin{equation}
\ell_i^B(x) =  \ell_{j}^T(x) \ell_{k}^T(x)
\label{eqn: like-spec}
\end{equation}
for all $x \in [0,1]$.

\subsubsection{Partial likelihoods along a branch}

Let $N_i$ be the effective population size for the branch directly above node $i$, and let $\tau_i$ denote the length of the branch above node $i$, measured in numbers of generations. Then $t_i = \frac{\tau_i}{2N_i}$ equals the length of the branch in coalescent units. To derive the partial likelihood $\ell_i^T(x)$ at the top of this branch, we integrate over all frequency values at the bottom of the branch. 
\begin{equation}
\ell_i^T(x) = \int_0^1 f(y,t_i) \ell_i^B(y) dy.
\label{eqn: like-branch}
\end{equation}
In this equation, $f(y,t_i)$ is the density for the allele frequency $y$ the bottom of the branch conditional on an initial value $x$ at the top of the branch, while  $\ell_i^B(y)$ is the partial likelihood conditional on allele frequency $y$ at the bottom of the branch.  

To get a handle on \eqref{eqn: like-branch} we define
\begin{equation}
	g(x,t) = \int_0^1 f(y,t) \ell_i^B(y) dy , \text{ for $0<t<t_i$}.
\end{equation} 
Then $g$  satisfies the PDE, 
\begin{equation}
\frac{dg(x,t)}{dt} = (\beta_1(1-x)-\beta_2x)\frac{d}{dx}g(x,t) + \frac{1}{2}  x(1-x)\frac{d^2}{dx^2}g(x,t). 
\label{eqn: backward diffusion}
\end{equation}
with initial condition 
\begin{equation}
g(x,0) = \ell_i^B(x)\text{, for } x \in [0,1].
\label{bc: initial}
\end{equation}

Solutions to the PDE \eqref{eqn: backward diffusion} are unique and well-behaved without specifying boundary conditions for $t>0$ \citep{epstein2010wright}. 
\subsubsection{Likelihood at the root of a tree}

Equations~\eqref{eqn: like-tip}, \eqref{eqn: like-spec} and \eqref{eqn: like-branch} can be applied to the entire tree, starting with the leaves and moving upwards towards the root. They define the root likelihood $\ell_\rho^B(x)$, which is the probability of observing all allele counts for all populations, conditioning on the allele frequency $X_\rho = X_\rho^B$ being equal to $x$. 

To define the likelihood of the tree we now integrate over the allele frequency $x$ at the root, multiplied by the stationary density of $x$ given in \eqref{eqn: stationary distr.}:
\begin{equation}
L_m = \int_{0}^{1} \pi(x|\beta_1, \beta_2) \ell_{\rho}^B(x) dx.
\label{eqn: likelihood}
\end{equation}
This is the probability for a single site $m$. The likelihood from all markers is found by multiplying the probabilities from each site together (or, in practice, summing the log-likelihoods) 
\begin{equation}
	 \log L = \sum_{m} \log(L_m).
\end{equation}

In the case of SNP data we need to compute the conditional likelihood, $L_m/(1- L_0)$, where $L_0$ is the probability of observing a constant site (See \citet{Bryant2012} and \citet{felsenstein1992phylogenies}).

\subsection{Translation of parameters}

One of the more confusing aspects of working across population genetics and phylogenetics is the different ways that different communities parameterize different variables. In this section we summarise the parameters and outputs for our model, and show how they might be converted into other formulations. 

There are four groups of input variables in our model. These are the variables that could conceivably be inferred from data using the likelihood function. They are
\begin{enumerate}
\item The species tree. 
\item The branch lengths in the species tree. In our model, we measure the length of a branch in {\em number of generations}. Let $\tau_i$ denote the length of the branch above node $i$.
\item The mutation rates $u$ and $v$ giving the probabilities of mutating from a red to a green allele or a green to a red allele, {\em each generation}. 
\item The effective population size $N_i$ for each branch $i$ (that is, the branch above node $i$). 
\end{enumerate}

Different methods for inferring species trees describe these parameters in different ways. The convention in phylogenetics is to express branch lengths in terms of { \em expected substitutions per site}. The use of the term 'substitution' is confusing, as the whole premise behind the multispecies coalescent is to model incomplete substitutions. However for neutral mutations, the rate of substitutions equals the rate of mutations. Under our model, the rate of mutations per generation is 
\begin{align}
 \mu &= P(\mbox{ lineage has red allele }) u + P(\mbox{ lineage has green allele }) v \\
     &= \frac{v}{u+v} u + \frac{u}{u+v} v,\label{muform} 
\end{align}
so a branch of length $\tau_i$ generations corresponds to a branch length of $\mu \tau_i$ expected substitutions (mutations) per site. 

Methods which ignore the branch lengths in gene trees are not able to infer both population sizes and branch lengths, and instead use a single parameter per branch, typically measured in coalescent units (e.g. \citep{vachaspati2015astrid,liu2010maximum}). If $\tau_i$ is the length of a branch in numbers of generations,  the length of the branch in coalescent units is given by $t_i = \frac{\tau_i}{2N_i}$. 

In our model, we parameterize effective population size for branch $i$ directly as $N_i$. \snapp \citep{Bryant2012}, Best \citep{liu2008best} and BPP \citep{yang2015bpp} instead use $\theta_i = 4 N_i \mu$ as the parameter for effective population size. Under an infinite sites model, $\theta_i$ equals the expected proportion of sequence differences between two individuals from the same population. In a finite sites model, $4 N_i \mu$ is an overestimate of this expectation, due to backward mutations. 

\subsection{Dealing with confounded rates and rate matrices}

An important issue with the diffusion model as we have described it, and indeed with many multispecies coalescent models, is that there is an identifiability problem with rates. If the mutation rates are multiplied by some constant $c$ and at the same time branch lengths and population sizes are multiplied by $\frac{1}{c}$, the probability of the data remains the same. 

One solution is to estimate the mutation rate $\mu$ beforehand and use this value, or a prior distribution around that value, to include $\mu$ as part of our model. The prior distribution for $u$ and $v$ is then reformulated so that \eqref{muform} is satisfied. This strategy is used by StarBeast, where the average substitution (mutation) rate $r_\mu$ is fixed ahead of time. 

An alternative strategy is to express branch lengths in terms of expected substitutions (mutations) per site and population sizes in terms of the $\theta$ parameter, as both of these are invariant to a choice of $\mu$, \citep{yang2015bpp,Bryant2012}. As a consequence, the effective population sizes $N_i$ cannot be inferred without additional information. This approach adapts well to phylogenetics where it is customary to describe mutation rates using a normalised rate matrix, such as 
 the Jukes-Cantor and HKY85 matrices
\[\mathbf{Q} = \left[ \begin{matrix} - & 1/3 & 1/3 & 1/3 \\1/3  & - & 1/3 & 1/3 \\ 1/3 & 1/3 & - & 1/3 \\ 1/3 & 1/3 & 1/3 & - \end{matrix} \right] \mbox{ and } \mathbf{Q} = \frac{1}{r}\left[ \begin{matrix} - & \pi_C & \kappa \pi_G & \pi_T \\ \pi_A & - & \pi_G & \kappa \pi_T \\ \kappa \pi_A & \pi_C & - & \pi_T \\ \pi_A & \kappa \pi_C & \pi_G & - \end{matrix} \right].\]
In this matrix, the stationary probabilities for the nucleotides are $\pi_A,\pi_C,\pi_G,\pi_T$, the parameter $\kappa$ controls the ratio of transitions to transversions, and the diagonal elements are chosen to make rows sum to zero. The Jukes-Cantor matrix on the left is already normalised, while in the HKY85 model $r$ would be chosen to make the overall substitution rate equal to $1$.

These matrices can be incorporated directly into \snapper, either using prior information for $\mu$, or when using the same branch length and population parameter scheme as BPP. For any site, if $X$ and $Y$ are the two nucleotides present and we (arbitrarily) assign red to $X$ and green to $Y$ then the appropriate choices for $u$ and $v$ are simply $\mathbf{Q}_{XY}$ and $\mathbf{Q}_{YX}$. For example, under the HKY85 model,  if the red allele is nucleotide A and green allele is nucleotide C then the corresponding rates are $u = \mu \mathbf{Q}_{AC} = \pi_C$ and $v = \mathbf{Q}_{CA} \pi_A$. We are approximating a four state model with a two state model, but the error introduced should be minimal in the absence of divergences or highly variable sites.

\subsection{Comparison with allele frequency spectrum methods}

We note that diffusion models are already used in population genetics to approximate changes in the {\em allele frequency spectrum} (AFS) \citep{Gutenkunst2009, Hey2012, racimo2016joint}. For each value of $x$ between $0$ and $1$, the AFS gives the proportion of sites for which the derived allele appears in the proportion $x$ of the sample. For example, we might consider all sites where the the derived allele appears in 30 of the 100 sampled genomes. If $5\%$ of all sites fall into this category then the AFS value for 0.3 will be $5\%$.

Methods based on the AFS to infer genetic parameters for a single population are widespread in the literature \citep{boitard2016inferring, lapierre2017accuracy, nielsen2000estimation}. In all of these analyses, it is assumed that the population dynamics have achieved stationarity when the AFS predicted by the model is used. 

When there are multiple populations divergence times become important. Suppose that there are $s$ species. For every vector $(x_1,x_2,\ldots,x_s)$ of numbers between $0$ and $1$, the joint-AFS gives the proportion of sites where the derived allele appears in a proportion $x_1$ of the individuals in the first population, $x_2$ of the individuals in the second population, and so on.  While the AFS for each population by itself will be the stationary AFS, the joint-AFS for multiple populations depend on the time since separation. Fortunately the PDE determining the predicted joint-AFS is a straight-forward extension of the PDE for a single population \citep{Gutenkunst2009, Hey2012, racimo2016joint}.

There are two main advantages of an approach based on the joint-AFS when compared to the method we describe here. First, the PDE for the joint-AFS only needs to be solved once for all sites, whereas in our approach we end up solving the PDEs once for each (distinct) site. Second, migration between populations can be incorporated quite simply into AFS calculation, whereas it would break down the dynamical programming strategy that we use.

The main disadvantage of the joint-AFS strategy is that the PDE has as many dimensions as the number of species, and suffers from the {\em curse of dimensionality} \citep{Gutenkunst2009}, leading to an exponential growth in the size of grid used by standard numerical methods. This factor severely limits the number of species which can be considered concurrently, whereas in our approach the algorithm scales linearly with the number of species.

\section{Computing the likelihood efficiently}

In the previous section, we showed how the recursions for the partial likelihoods can be derived analytically. We did not show how to actually compute those likelihoods. Indeed, computing the likelihoods amounts to an extremely high dimensional integration problem. Our approach is to use numerical techniques combined with dynamic programming. While the likelihoods we compute are approximate, we can still bound the error. 

\subsection{Shifted Chebyshev polynomials}

\citet{Hiscott2016} describe a general strategy for computing likelihoods numerically whereby partial likelihoods are evaluated on a mesh of values at each node, and accurate quadrature methods used to carry out the actual computation. We will extend the same strategy by using a set of basis functions to express approximate partial likelihoods instead of a mesh of values. The basis functions are cleverly chosen to help solve equation \eqref{eqn: backward diffusion} efficiently and accurately.

The basis functions we use are called the {\em shifted Chebyshev polynomials of the first kind}, denoted \[T_0^*(x),T_1^*(x),T^*_2(x),\ldots.\] 
Shifted Chebyshev polynomials are defined on $[0,1]$ and have  particularly nice properties \citep{FoxParker1968}. 
They also have a lot of equivalent definitions, but the simplest uses the following recursion, 
\begin{align*}
	T^*_0(x) &= 1 \\ 
	T^*_1(x) &= 2x-1 \\ 
	T^*_k(x) &= 2(2x-1)T^*_{k-1}(x) - T^*_{k-2}(x).
\end{align*}
The shifted Chebyshev polynomials are related to the better-known Chebyshev polynomials $T_0,T_1,T_2,\ldots$ by the identity:
$T^*_k(x) = T_k(2x -1)$.
That is, they are obtained by shifting and scaling the Chebyshev polynomials to have domain $[0,1]$, see \citet{MasonHandscomb2003}.

There are two main ways of using shifted Chebyshev polynomials to approximate  $\ell_i^B$ and $\ell_i^T$ as functions of $x$. The first is to approximate the function directly as a linear combination of the shifted Chebyshev polynomials, that is, finding sets of coefficients $\lambda^B_{i,0},\lambda^B_{i,1},\lambda^B_{i,2},\ldots$ and $\lambda^T_{i,0},\lambda^T_{i,1},\lambda^T_{i,2},\ldots$ so that for all $x$ in $[0,1]$ we have
\[ \ell_i^B(x)\approx \sum_{k=0}^{K}\lambda_{i,k}^BT_k^*(x)  \quad \text{and} \quad \ell_i^T(x) \approx \sum_{k=0}^{K}\lambda_{i,k}^TT_k^*(x) .\]
It can be shown that the error in this approximation drops exponentially quickly as the number $K$ of basis functions increases. We therefore only need to store a few coefficients in order to evaluate the partial likelihood function at any $x$, with small error. 

The second way of obtaining an approximation is by determining the values $\ell_i^B(x)$ and $\ell_i^T(x)$ at a pre-specified set of points $x_0,x_1,\ldots,x_K$ and then finding the unique combination of coefficients $\lambda_{i,0}^B,  \dots  ,\lambda_{i,K}^B$ such that
\[\sum_{k=0}^K \lambda_{i,k}^B(x_j) = \ell_i^B(x_j)\]
for all $j=0,1,\ldots,K$. This is called polynomial interpolation. As it happens, there is a particular choice of points $x_0,\ldots,x_K$  for which we can switch back and forth between function values 
\[\ell_i^B(x_0),  \dots  ,\ell_i^B(x_K) \]
and interpolation coefficients
\[\lambda_{i,0}^B,  \dots  ,\lambda_{i,K}^B \]
and back with little numerical error and in $O(K \log K)$ time \citep{waldvogel2006,trefethen2013approximation}. These are called the Chebyshev-Lobatto points, defined by the rather opaque formula
\[x_j = \left(1-\cos\left(\frac{2 \pi j}{K}\right)\right)/2, \text{ for $j =0,\dots,K$} .  \] 
These points are all in the interval $[0,1]$ with a denser packing of points nearer $0$ and $1$. They equal the $x$-coordinates of points spaced regularly around a semi-circle. 

We use both coefficient values $\lambda_{i,k}^B,\lambda_{i,k}^T$ and function values $\ell_{i}^B(x_j),\ell_{i}^T(x_j)$ when approximating the partial likelihoods $\ell_i^B$ and $\ell_i^T$. 

\subsection{Approximate partial likelihoods at a leaf}

We compute our approximate likelihood at the bottom of a leaf branch by simply evaluating the function 
\begin{equation}
\ell_i^B(x_j) = {2n_i \choose r_i} x_j^{r_i}(1-x_j)^{2n_i - r_i} \text{ for j = 1, \dots , K}
\label{eqn: approx-like-tip}
\end{equation}
at the Gauss-Lobatto points.
We then compute corresponding coefficients $\lambda_{i,0}^B,\lambda_{i,1}^B,\ldots,\lambda_{i,K}^B$ using the FFT algorithm \citep{trefethen2013approximation}. 

\subsection{Solving the diffusion model numerically}

Shifted Chebyshev polynomials provide the foundation for the numerical methods we use to solve the PDE \eqref{eqn: backward diffusion}. Note that, unlike the forward diffusion \eqref{eq:forward}, solutions to this backward diffusion are nice smooth functions, meaning that we can avoid some of the numerical headaches encountered by those using the forward diffusion directly \citep{Hey2012}.

We approximate $g(x,t)$ in terms of shifted Chebyshev polynomials
\begin{equation}
g(x,t) \approx \sum_{k=0}^{K}\lambda_k(t)T^*_k(x).
\label{eqn: approx_g}
\end{equation}
Then 
\begin{align}
\frac{\partial g(x,t)}{\partial t}  &\approx \sum_{k=0}^K \frac{d\lambda_k(t)}{dt} T^*_k(x),  \label{eqn: dg/dt} \\
\frac{\partial g(x,t)}{\partial x}  &\approx \sum_{k=0}^K \lambda_k(t)    \frac{d T^*_k(x) }{dx} 
\intertext{and}
\frac{\partial^2 g(x,t)}{\partial x^2}  &\approx \sum_{k=0}^K \lambda_k(t)    \frac{d^2 T^*_k(x) }{dx^2}.
\end{align}
Let ${ \bm \lambda}(t)$ denote the vector of values $\lambda_0(t),\ldots,\lambda_K(t)$.  Formulas for the derivative of shifted Chebyshev polynomials lead eventually to a $(K+1) \times (K+1)$ matrix $\bm{D}$ such that 
\[ \sum_{k=0}^K \lambda_k(t)    \frac{d T^*_k(x) }{dx}  =  \sum_{k=0}^K (\bm{D} \bm{\lambda}(t))_k    T^*_k(x),\]
and
\[ \sum_{k=0}^K \lambda_k(t)    \frac{d^2 T^*_k(x) }{dx^2}  =  \sum_{k=0}^K (\bm{D}^2 \bm{\lambda}(t))_k    T^*_k(x).\]
Indeed after some tedious algebra we can derive a $(K+1) \times (K+1)$ $\bm{Q}$ matrix so that
\[ (\beta_1(1-x)-\beta_2x)\frac{d}{dx}\sum_{k=0}^{K}\lambda_k(t)T^*_k(x) + \frac{1}{2}  x(1-x)\frac{d^2}{dx^2}\sum_{k=0}^{K}\lambda_k(t)T^*_k(x) =  \sum_{k=0}^K (\bm{Q} \bm{\lambda}(t))_k    T^*_k(x)\]
for all $x \in [0,1]$. Plugging this and \eqref{eqn: dg/dt} into \eqref{eqn: backward diffusion} we obtain an equation  
\[\sum_{k=0}^K \frac{d\lambda_k(t)}{dt} T^*_k(x)  =  \sum_{k=0}^K (\bm{Q} \bm{\lambda}(t))_k    T^*_k(x)\]
for an approximate solution to the PDE, giving the system
\begin{equation}
	\frac{d{\bm \lambda}(t)}{dt}  = {\bm Q} \cdot {\bm \lambda}(t).
	\label{eqn: approx_dt}
\end{equation}
with initial condition ${\bm \lambda}(0)$ given as the Chebyshev coefficients of $g(x,0)$.

Like \citet{Bryant2012} we use rational approximations to $\exp({\bf Q}t)  \bm{\lambda}(0) $ on the negative real axis computed using a Caratheodory Fejer-approximation. Additionally, we use techniques adapted from \citet{FoxParker1968} and outlined in the appendix (A.1) to take advantage of structure in the matrix $\mathbf{Q}$, allowing an implementation of the 
 Caratheodory-Fejer approximation which runs in $\mathcal{O}(K)$ time.
 
To summarise, consider a node $i$ and let $t_i$ denote the length (in coalescent units) of the branch connecting $i$ to its parent. Suppose $\ell_i^B$ with corresponding coefficients $\lambda^B_i$  is already computed at the bottom of the branch. We then compute partial likelihood $\ell_i^T$ at the top of the branch by
\begin{enumerate}
	\item Setting ${\bm \lambda}(0) = \lambda^B_{i,0},\dots, \lambda^B_{i,K}$
	\item Computing a numerical approximation for $\exp({\bm Q} t_v) {\bm \lambda}(0)$. 
	\item $\ell_i^T$ = $\lambda(t_0)_0,\dots, \lambda(t_0)_K$
\end{enumerate}

\subsection{Approximate partial likelihoods at a speciation}

	Consider a node $i$ with two children $j$ and $k$. We suppose that the approximations for the partial likelihoods $\ell_{j}^T(x_0),\ell_{j}^T(x_1),\ldots \ell_{j}^T(x_K)$ and $\ell_{k}^T(x_0),\ell_{k}^T(x_1),\ldots \ell_{k}^T(x_K)$. We then have 
	\begin{align}
	\ell_i^B(x_p)= \ell_{j}^T(x_p) \ell_{k}^T(x_p) & \mbox{ for all $p=0,1,\ldots,K$.}
	\label{eqn: approx-species}
	\end{align}
	 This computation takes $O(K)$ time.

\subsection{Approximate likelihoods at the root} 

The final integration to carry out is over the partial likelihood at the root
\[ \int_{0}^{1} \pi(x|\beta_1, \beta_2) \ell_{\rho}^B(x) dx,\]
see \eqref{eqn: likelihood}. The density for a beta distribution 
\eqref{eqn: stationary distr.} can be infinite at the boundaries, meaning that numerical integration techniques such as Clenshaw-Curtis quadrature can give poor approximations. The solution is to separate out those parts which are difficult to integrate numerically and determine them analytically. Suppose we have an approximation of $\ell_{\rho}^B(x)$ as a polynomial
\[f(x) =  \sum_{k=0}^{K} \lambda_{\rho,k}^B T^*_k(x) \approx \ell_\rho^B(x)  .\]
We factor this polynomial as 
\[ f(x)  = x(1-x) g(x) + f(0) + (f(1)-f(0))x\]
and then compute
\[ \int_0^1 f(x)  \pi(x|\beta_1, \beta_2) dx = \int_0^1 g(x) x(1-x) \pi(x|\beta_1, \beta_2) dx + \int_0^1 (  f(0) + (f(1)-f(0))x)  \pi(x|\beta_1, \beta_2) dx. \]
Noting that the first integral is now well-behaved while the second integral  evaluates to $f(0) + (f(1)-f(0))\frac{\beta_1}{\beta_1 + \beta_2}$ by properties of the Beta distribution.

\subsection{Computing the log-likelihood of a species trees}

Algorithm \ref{algorithm: compute_likelihood} summarises the numerical calculation of the likelihood. This algorithm takes $\mathcal{O}(sKlog(K))$ time per site where $\mathcal{O}(ns)$ pre-processing, where $s$ is the number of species, $K$ the number of Chebyshev basis functions and $n$ is the number of individuals at a site. The pre-processing step involves counting the frequencies of allele types in each population. In practice this step could be carried out once per data set, rather than once per tree evaluated.   

\begin{algorithm}
	\caption{\snapper: Computes tree log-likelihoods}\label{alg:generate_fake_data}
	\begin{algorithmic}[1]
		\Procedure{log-likelihood}{$X$, S}
		\For{All sites $s$}
		\For{All nodes $i$ in a post-order traversal of the tree}
		\If{$i$ is a leaf node}
			\State{Evaluate $\ell_i^B(x_j)$ for each $j$ using \eqref{eqn: approx-like-tip}}
		\Else
			\State{Evaluate $\ell_i^B(x_j)$ for each $j$ using \eqref{eqn: approx-species}}
		\EndIf
		\State{Compute coefficients $\lambda_{i,j}^B$ from values $\ell_i^B(x_j)$}
		\State {Compute coefficients $\lambda_{i,j}^T$ by solving the PDE \eqref{eqn: approx_dt} with ${\bm \lambda}(0)= {\bm \lambda}_{i}^B$}
		\State{Compute values $\ell_i^T(x_j)$ from coefficients $\lambda_{i,j}^T$}
		\EndFor
		\State{Compute $L_s$ using numerical integration.}
		\EndFor
		\State \textbf{return} $\sum_s log(L_s)$		
		\EndProcedure 
	\end{algorithmic}
	\label{algorithm: compute_likelihood}
\end{algorithm}

The conversion to and from coefficients to function values in the Chebyshev expansion in lines 7 and 9 each takes $O(K \log K)$ time, using the FFT algorithm reviewed above. Solution of the PDE in line 10 takes $O(K)$ time. 
		
\subsection{\snapper}

The likelihood algorithm forms the core of a Bayesian inference software package, \snapper. The software is open-source and available to download at \url{https://github.com/rbouckaert/snapper}. Like \snapp, it takes biallelic data at multiple loci for multiple individuals, in a set of species and returns samples from the joint posterior distribution of (i) species phylogenies, (ii) species divergence times and (iii) effective population sizes. As in  \snapp we implemented a dynamic cache-based system to store partial likelihoods on different subtrees and multithreading to take advantage of parallel computation on multiple core  machines or graphics processing units. The range of prior distributions,  and flexible prior specification, remains essentially unchanged. 

The MCMC proposal function implemented is almost the same subset of BEAST2 \citep{bouckaert2019beast2}
move proposals implemented for \snapp. We add one new proposal selects a subtree of the species tree and scales all population sizes within that subtree simultaneously, a refinement of an existing proposal.

The user can specify number $K$ of basis functions.
Some recommendations are made in the \snapper manual regarding the number of basis functions to use given the size of a dataset. Also contained in the software package are simulators and python scripts to integrate \snapper with the iPyRad data pipeline to assist with streamlining data analysis.

\section{Performance assessment}

We have tested the likelihood algorithm and the \snapper software extensively. Here we highlight some results related to the performance of \snapper. Details of additional simulations, including tests checking that posterior distributions recovered model parameters, can be found in the appendix (A.2). 

We conducted the following experiments: (i) An assessment of error convergence as the number $K$ of Chebyshev basis functions increase (ii) A comparison of inferred parameters between \snapp and  \snapper given a dataset (iii) An assessment of how computational time estimates for \snapp and \snapper scale in terms of size of the data set. All simulations and inference were run on a laptop with an Intel i7-8565U CPU.   

\subsection{Numerical error convergence}
 We assessed the rate of error convergence for different tree heights, tree sizes and tree topologies. In particular we wanted to verify that we have exponential convergence of error.      

\begin{figure}[h]
	\includegraphics[scale=0.5]{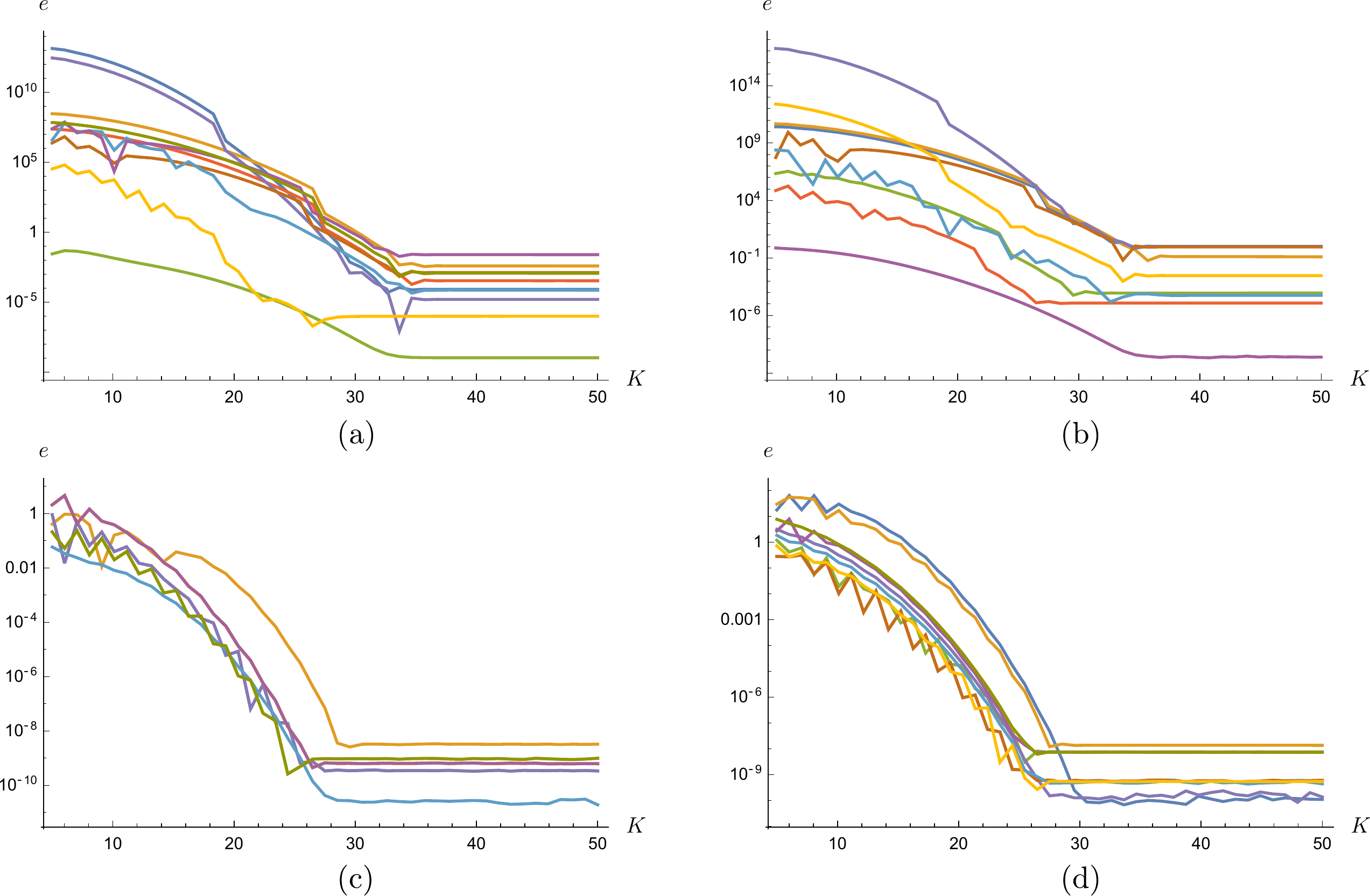}
	\caption{Log scale plot of relative error for basis functions $K=5,6,\dots,50$ of 4-taxa trees:  (a) balanced short; (b) caterpillar short; (c) balanced tall; (d) caterpillar tall. The sub-linear decrease in log-error corresponds to an exponential decrease in error, until the limit of machine precision is reached and no further improvements in error are possible. 
	 }
	\label{fig: numerics-4-taxa}
\end{figure}

\begin{figure}[h]
	\includegraphics[scale=0.5]{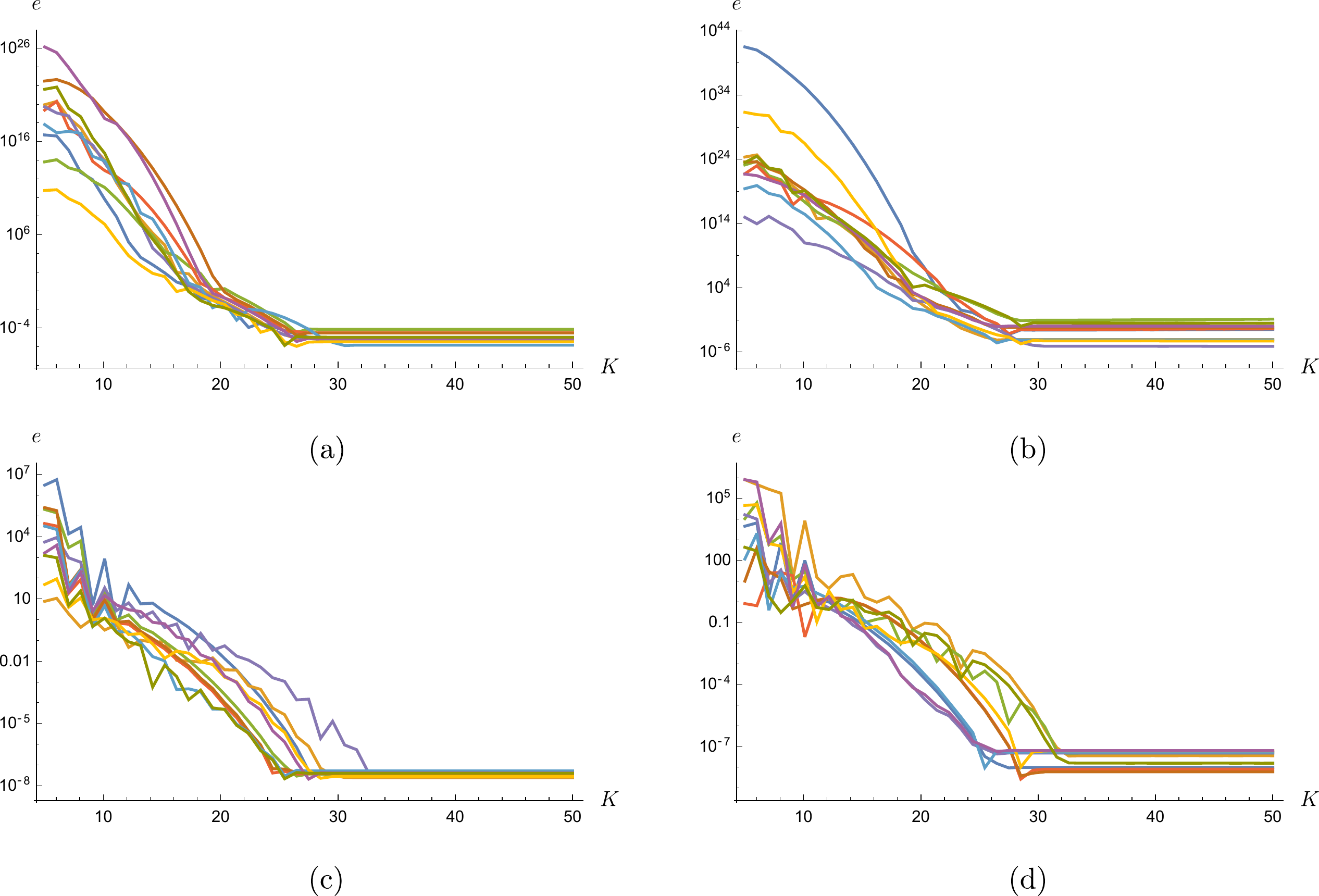}
	\caption{Log scale plot of relative error for basis functions $K=5,6,\dots,50$ of 16-taxa trees: (a) balanced short; (b) caterpillar short; (c) balanced tall; (d) caterpillar tall. The sub-linear decrease in log-error corresponds to an exponential decrease in error, until the limit of machine precision is reached and no further improvements in error are possible. 
}
	\label{fig: numerics-16-taxa}
\end{figure}

We drew parameter sets from  the prior distributions specified by Table \ref{tbl: modelparameters_gen}, for 4 species trees and 16 species trees with `caterpillar' and `well-balanced' tree shapes. Furthermore for each tree we specified expected tree heights to be either very small or very large. We then simulated a 1000 sites for each parameter set and computed the relative error of the likelihood for number of basis function, $K=5,\ldots,50$. Note that for this experiment we do not limit $K$ to values of the form $2^m+1$. As there is no analytical expression for the likelihood so  to assess convergence we compared values calculated to those for a large ($K=200$) number of basis functions. We summarize the results for 4 species trees in Figure \ref{fig: numerics-4-taxa} and for 16 species trees in Figure \ref{fig: numerics-16-taxa}. 

\begin{table}
	\begin{tabular}{clcccc}	
		\hline	
		& & \multicolumn{2}{c}{Short trees} & \multicolumn{2}{c}{Tall trees}\\
		Parameter & Description & Distribution & Mean & Distribution & Mean\\
		\hline 
		$\theta$ & Population size & $\Gamma$(2, 200) & 0.01 & $\Gamma$(2, 200) & 0.01  \\
		Tree & Shape and branch length & Yule(100) &  0.00111... &  Yule(0.1) & 10
	\end{tabular}
	\caption{Priors used to draw trees for error convergence study. Gamma priors on population sizes and Yule prior on tree shape, height and branch length was specified.}
	\label{tbl: modelparameters_gen}
\end{table}

We see that in all the cases we have that the error decreases exponentially with $K$, that is, it decreases like $\alpha^K$ for some $\alpha<1$. Error is smaller for long branch lengths since approximate solutions are typically lower degree polynomials. However when the branch lengths are
very short the error decreases more slowly. There are good reasons for this. Firstly, when the partial likelihood function is spiked, the approximate solutions are high degree polynomials requiring more basis functions. Secondly, population sizes for short branches are intrinsically more difficulty to estimate, no matter which model or method is used. The only information we have about population sizes comes from the distribution of coalescent events, and on short branches there are simply insufficient coalescent events to make sound inference. Later, we address this by adopting a prior distribution on population sizes which introduces correlation between neighbouring branches.

Apart from branch length distribution, tree shape does not seem to have a noticeable effect on the error convergence. 

\subsection{Comparing inferences from \snapp and \snapper}

Theory states that the multispecies coalescent model underlying \snapp and the diffusion model behind \snapper are approximations of one another \citep{Griffiths2010}, with differences decreasing as effective population size increases. To demonstrate this we analyze the rattlesnake data set in \citet{chifman2014quartet}. The data 3000 SNPs from a total of 59 individuals in  7 populations. 

In \citet{chifman2014quartet} it was reported that \snapp was not sampling from the posterior distribution efficiently enough. It quickly became apparent that the sole cause of difficulty was the inference of population sizes on branches with close to zero length. We selected the CIR prior in \snapp and in \snapper, which implements a model with correlated population  sizes on nearby branches. Convergence was also improved by implementing a proposal which scales all population sizes within a randomly selected subtree.

We ran the MCMC chain for 750,000 iterations for both \snapp and \snapper. It took \snapp approximately $\sim 56$ hours to run and \snapper  approximately $\sim 4$ hours to run. We give the summary statistics in the appendix (A.3). 

Figure \ref{fig: 7taxon} uses Densitree \citep{bouckaert2010densitree} to depict samples of the posterior distributions of the most likely densitrees for \snapp and \snapper. We print the posterior mean of effective population size above each branch.  In both cases we see that there is only one topology in the 95\% highest posterior density. Posterior distributions of branch lengths are mostly indistinguishable. There are some small differences in posterior distributions of populations sizes. However in all cases population sizes follow the same apparent distributions. 

\begin{figure}[h]
	\centerline{
	\includegraphics[width=\textwidth]{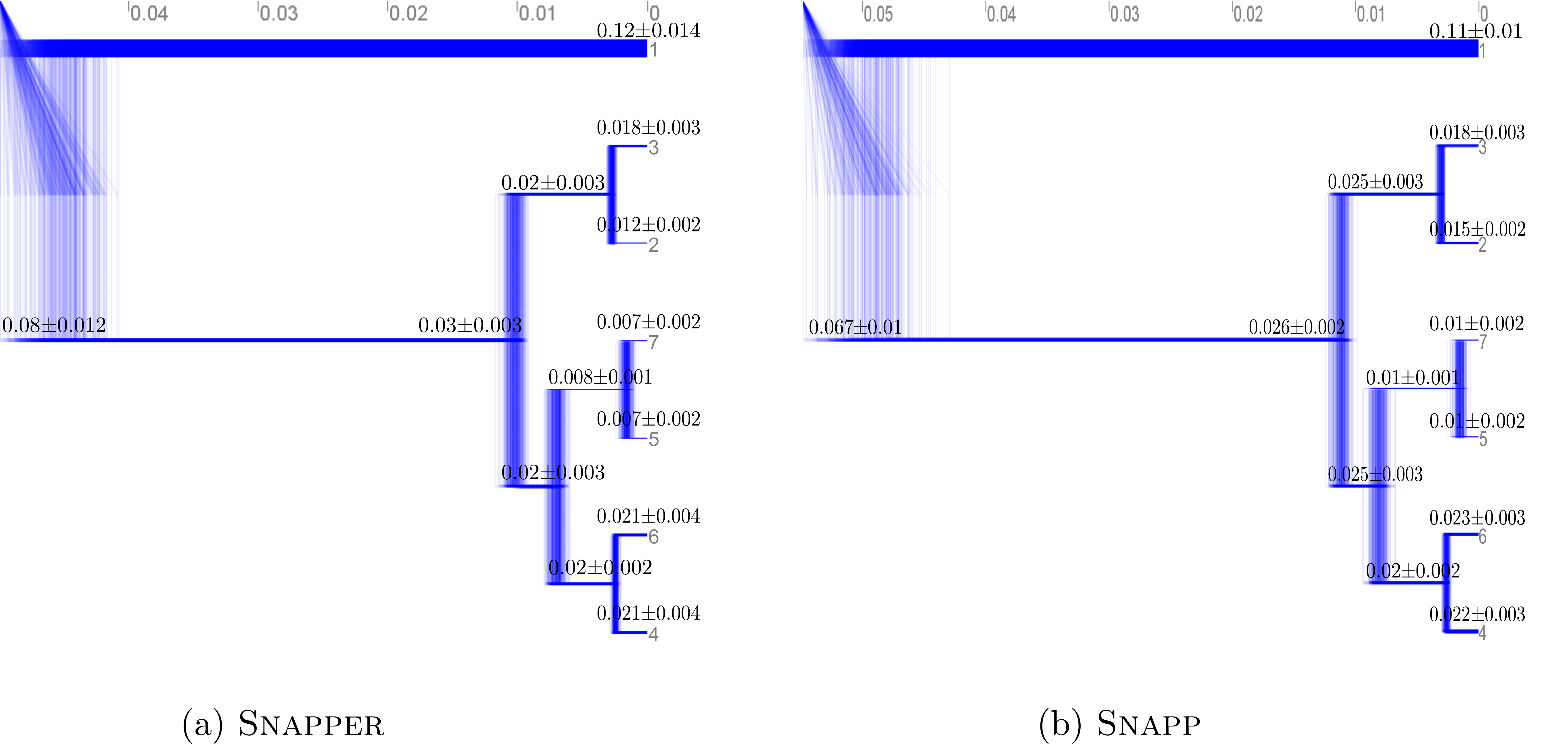}}
	\caption{A \snapp and \snapper inference of rattlesnake species trees displayed using Densitree \citep{bouckaert2010densitree}. Rattlesnake populations are numbered from 1-7. Branch thickness is related to relative population sizes, tree height is reported in expected number of mutations and population sizes is printed below each related branch, as $\mu \pm 3\sigma$ }
	\label{fig: 7taxon}
\end{figure}

\subsection{Running time of \snapper}

\begin{figure}
	\centering
	\includegraphics[scale = 0.4]{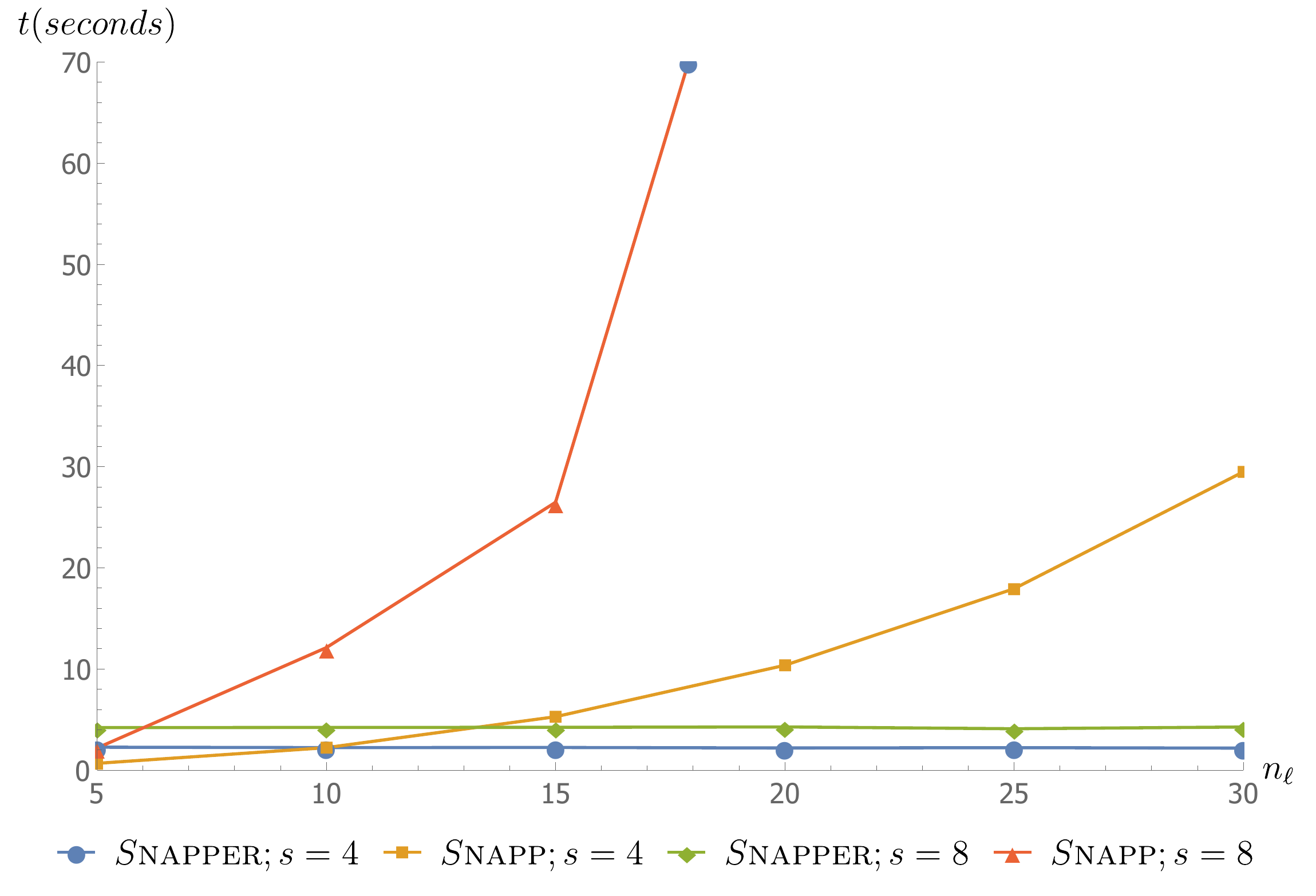}
	\caption{ Average times (in seconds) to compute the likelihood of a 1000 sites on 4- and 8-taxa trees, for \snapp and \snapper with $s=4,8$ (number of taxa) and $n_{\ell} =5,10,\dots,30$ (number of individuals at each leaf).} 
		\label{fig: comp_time}
	\end{figure}

We compare the computation time between \snapper and \snapp on a 4- and 8-taxa tree scaling the number of individuals at a leaf.
We generated datasets of a 1000 SNPs given the number of species ($s=4,8$) and the number of individuals at a leaf ($n_{\ell} =5,10,\dots,30$). The number of basis functions for \snapper was fixed at K = 33. Likelihood computation without caching was done on a computer with an Intel i3-7100 CPU. 
In Figure \ref{fig: comp_time} we report the average times (in seconds) to compute the likelihood of a 1000 sites on 4- and 8-taxa trees, for \snapp and \snapper. As expected we can clearly see the advantage of \snapper, in that computation time is hardly affected by the number of individuals at a leaf. Note that there is some dependence on the number of individuals. As having large samples can lead to a need for more basis functions to ensure accuracy. 

\section{Analysis of freshwater turtle; Emydura macquarii}
To illustrate the application of \snapper we reanalyse SNP data of \citet{georges2018genomewide} from a group of freshwater turtles known collectively as Emydura. The range of Emydura extends almost the full length of the Australian continent from north to south. The group is currently recognized as a complex of closely related and morphologically distinct allopatric forms, variously regarded as species, subspecies or distinct morphological lineages \citep{georges2018genomewide}. The individual samples were taken from 57 distinct water bodies to study and clarify some of the evolutionary relationships among different Emydura populations found in Eastern Australia river basins and along the coast. The sampling covers the coastal drainages of eastern Australia, from the Hunter River in the south (New South Wales) to the Normanby River (Queensland) in the north; the rivers	of the Murray-Darling Basin (MDB), including the Paroo drainage, and the Lake Eyre Basin (LEB), and the intervening Bulloo Basin \citep{georges2018genomewide}.

For the \snapper analysis we used 399 individuals divided into 41 populations with a total of 5,186 filtered SNPs. We expect short branch lengths for some of the populations due to geographical proximity of sampling sites. Therefore we specify a CIR prior that allows for correlation between populations on nearby branches. The analysis include two out-taxon populations therefore we expect tree height to be large. Hence we use a Yule prior with appropriate mean tree height. We have no prior information on mutation rates therefore we fix $u=v=1$.  See Table \ref{tbl: priors_turtles} for more details.    

\begin{table}
	\centering
	\begin{tabular}{clcc}	
		\hline	
		Parameter & Description & Distribution & Mean\\
		\hline 
		$\theta$ & Population size & CIR(2, 200,1) & 0.01   \\
		Tree & Shape and branch length & Yule(2) &  0.2 
	\end{tabular}
	\caption{Priors used in \snapper analysis of freshwater turtle; Emydura macquarii. CIR priors on population sizes and Yule prior on tree shape, height and branch length was specified. Note that the mean given for CIR prior is the mean of the stationary process.}
	\label{tbl: priors_turtles}
\end{table}

\begin{figure}
	\centering
	\includegraphics[scale = 0.6]{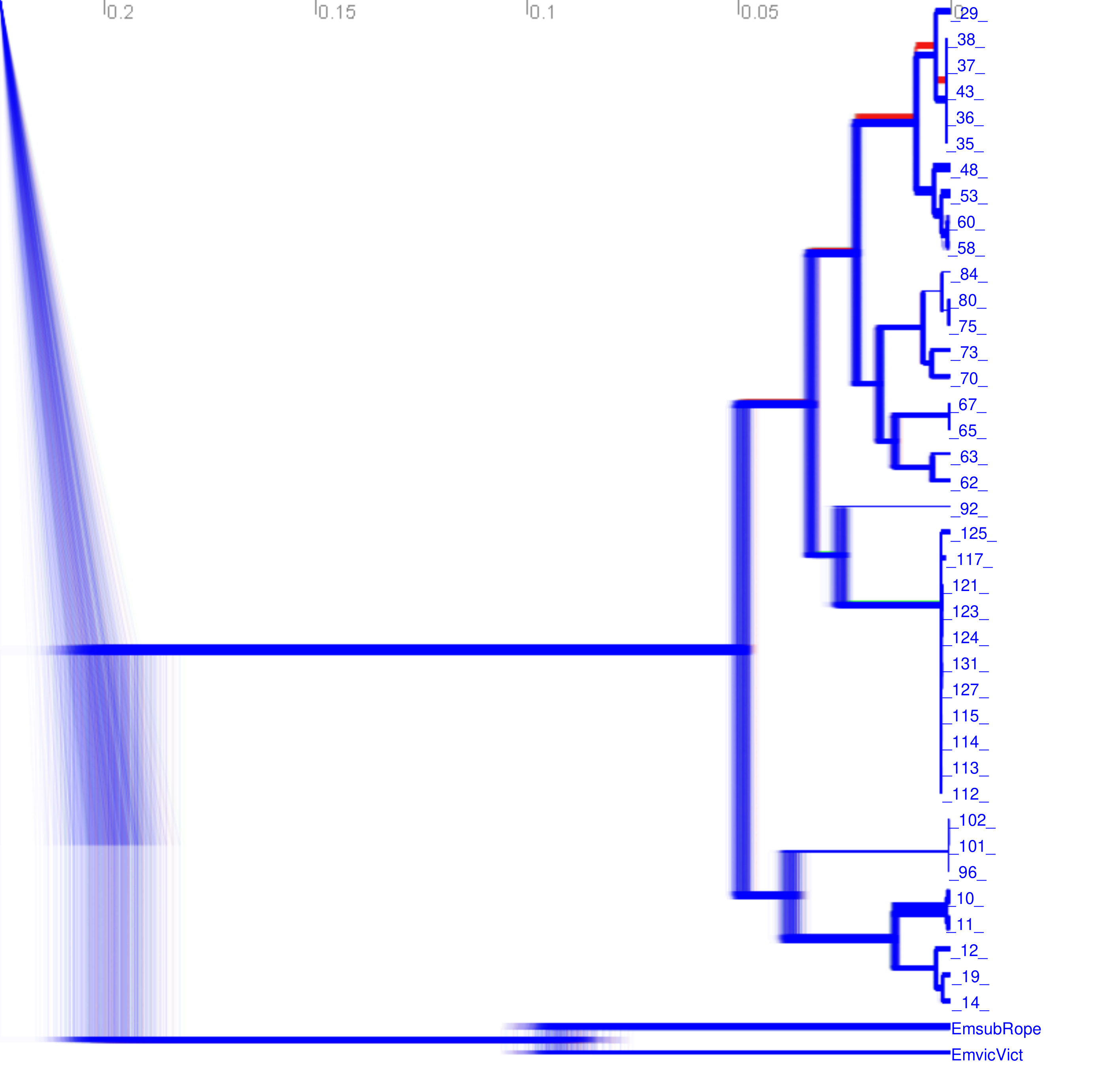}
	\caption{ 41 population densitree of freshwater turtle Emydura macquarii.
		On the x-axis, variation in the tree represents uncertainty of branch lengths. Thickness of the branches represent posterior mean of population sizes. Timescale grid at the top is given in expected number of mutations per lineage per site. }
	\label{fig: turtle_tree}
\end{figure}

We conducted a number of initial investigations to determine appropriate proposals and proposal weights. 
We then ran the sampler for 2,000,000 iterations with convergence assessed in Tracer \citep{rambaut2018posterior} using estimated effective sample size. It took a total of $\sim 500$ hours for the sampler to run on a computer with an Intel i3-7100. CPU. We provide a complete list of summary statistics in the appendix (A.3).

Figure \ref{fig: turtle_tree} depicts the tree with largest posterior probability with uncertainty. The shape of the tree in Figure \ref{fig: turtle_tree} agrees with the genetic distances and SVDquartets trees computed in \citet{georges2018genomewide}. We extend the anaylsis in \citet{georges2018genomewide} by quantifying uncertainty around branch lengths and include an additional parameter, namely population size. In Figure \ref{fig: turtle_tree} we see some uncertainty surrounding topology in the Fitzroy clade (samples 38 - 43). The Cooper Creek clade (samples 96 - 102) is sister to the North-east Coast clade (samples 10 - 19) in all three analyses, as is the sister relationship between the Hunter R (sample 92) and the Murray-Darling Basin (samples 96 - 115) populations. The East Coast (samples 29 - 60) and South-east Coast (samples 11 - 20) clade  is sister to the Hunter-MDB clade (samples 20 - 31) in the Snapper tree, in agreement with the Fitch-Margoliash distance tree, whereas the arrangement is unresolved in the SVDquartets tree. Sensibly, the Kolan-Burnett-Mary (sampels 48 - 58) clade is sister to the Fitzroy clade in the Snapper tree, whereas the SVDquartets tree has the Mary-Burnett clade in an unresolved trichotomy with the Cooper-NE Coast clade (samples 96 - 102; samples 10 - 19) and the Fitzroy clade. The \snapper analysis supports relationships within the NE Coast clade that reflect drainage proximity better than does the SVD quartets analysis.

\begin{figure}
	\centering
	\includegraphics[scale = 0.4]{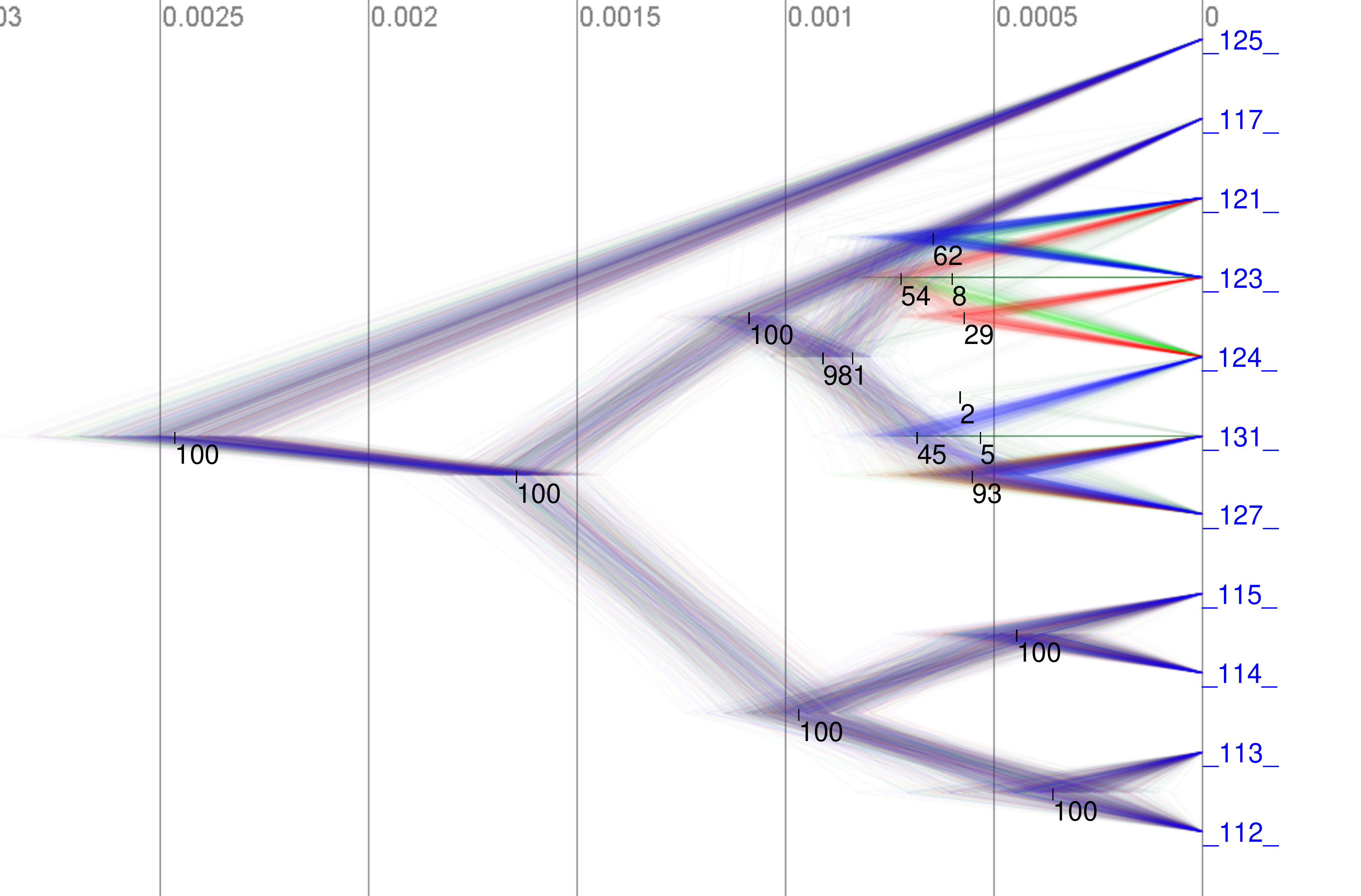}
	\caption{ 11 population densitree of freshwater turtle Emydura macquarii restricted to MDB clade to provide better resolution of branch lenghts and topology. We show the fraction of trees in the tree set that contain a clade as text on the graph. We also display the mean of a node as a marker on the tree. Timescale grid is given in expected number of mutations per lineage per site.  }
	\label{fig: turtle_tree_11}
\end{figure}

Also, we note that the bottom two population samples, i.e Emydura subglobosa  and Emydura victoriae populations, are considered as out-taxon groups. Therefore we expect to see the divergence time from the rest of the populations to be much earlier. This however leads to poor resolution for the MDB clade (samples 112 - 131). Thus we present the MDB clade in Figure \ref{fig: turtle_tree_11}.  The figure also depicts the extent of uncertainty surrounding the tree due to short branch lengths. As we discuss above, population size estimates here will be mostly dependent on the prior due to little information available on such short branch lengths. 

\section{Discussion}
In this paper we present \snapper a computationally more efficient method to supersede \snapp. Like \snapp the method takes biallelic markers 
sampled from individuals in multiple populations and computes the likelihood of the species tree topology together with branch lengths and population sizes. 
We achieved this computational efficiency by computing the likelihood using diffusion models rather than the multispecies coalescent. The Wright-Fisher diffusion and Coalescent are dual processes and it is therefore not surprising that the same inference can be made under these distinct but related model frameworks. 

We utilize observed allele frequencies in the sample as initial conditions for the backwards diffusion. The advantages of using the backward diffusion are two-fold. Firstly the numerical solutions are stable and bounded. Secondly, it is clear how to define the boundary conditions. 

The \snapper sampler is based on the \snapp sampler and uses the same move proposals, which are standard in the Beast2 software package. However to improve sampling efficiency for large trees we implemented an additional move to scale population sizes on subtrees.

We have reported some of the analyses performed in order to validate and more importantly convince the reader of the ability of \snapper to infer population genetic parameters. Further work needs to be done to establish regions in the appropriate parameter space where specific parameters are irrecoverable. 

Finally, we note that the general framework for defining likelihoods in terms of backward diffusions can be extended to other difffusion models. These models can include additional parameters such as selection, migration and linkage disequilibrium. In some cases even the computational framework is reusable.

\section*{Acknowledgements}
 We thank Arthur Georges for making the SNP dataset of Emydura macquarii available. Also for his help with the biogeographic interpretation of the Emydura macquarii species phylogeny and comparison of \snapper analysis with previous analysis done on the Emydura dataset. Marnus Stolz received a doctoral scholarship from the NZ Marsden Fund (PIs David Bryant and Steven Higgins).

 \bibliographystyle{sysbio}


  \appendix

\section{Appendix}
\subsection{The $O(K)$ algorithm for solving diffusions}

In the main text, we introduced a matrix $\mathbf{Q}$ which plays a key role in the numerical solution of diffusions. The matrix is chosen so that if 
\begin{equation}
g(x,t) = \sum_{k=0}^K \bm{\lambda}(t)_k T_k^*(x)
\label{eq:gexpand}
\end{equation} then 
\begin{equation}
(\beta_1(1-x)-\beta_2x)\frac{d}{dx}g(x,t) + \half  x(1-x)\frac{d^2}{dx^2}g(x,t) \approx \sum_{k=0}^K (\mathbf{Q} \bm{\lambda}(t))_k T_k^*(x). \label{eq:sparseOriginal}
\end{equation}
The bottleneck in the numerical method we use is the repeated solution of linear systems that look like 
\[(\mathbf{Q} - z \mathbf{I}) \mathbf{x} = \mathbf{v} \]
for difference complex values $z$ and vectors $\mathbf{v}$. Using a direct method, these take $O(K^2)$ time each as $\mathbf{Q}$ is lower triangular. However we can do better, using a trick described in \cite{FoxParker1968}. Here we give a very high level description of the approach.

The key idea is to apply integration twice to the LHS of \eqref{eq:sparseOriginal}. Using integration by parts multiple times we have
\begin{align}
\int_0^x \int_0^y \left[ (\beta_1(1-z)-\beta_2z)\frac{d}{dz}g(z,t) + \half  z(1-z)\frac{d^2}{dz^2}g(z,t) \right] \, dz\, dy  \hspace{-10cm} & \nonumber \\
& = - \int_0^x \int_0^y g(z,t) dz + \left( \beta_1(1-x) - \beta_2x - 1 + 2x \right)  \int_0^x g(z,t)dz \nonumber \\
& \quad \quad +  \half  x(1-x) g(x,t) + (\half - \beta_1)x g(0,t). \label{eq:doubleIntegrate}
\end{align}
We introduce two new matrices $\mathbf{X}$ and $\mathbf{Y}$. The matrix $\mathbf{X}$ is derived from properties of shifted Chebyshev polynomials, and is defined so that if $g$ is expanded as in \eqref{eq:gexpand} then 
\begin{align*} 
  \int_0^x \int_0^y  g(z) dz \, dy &\approx \sum_{k=0}^K (\mathbf{X} \bm{\lambda}(t))_k T^*_k(x) 
\end{align*}
The matrix $\mathbf{Y}$ comes from  \eqref{eq:doubleIntegrate} and has the property that if $g$ is expanded as in \eqref{eq:gexpand} then the RHS of \eqref{eq:doubleIntegrate} equals
\[  \sum_{k=0}^K (\mathbf{Y} \bm{\lambda}(t))_k T^*_k(x).\]
We then obtain 
\[ \mathbf{X} \mathbf{Q} \approx \mathbf{Y}.\]
The usefulness of this follows from that fact that, with the exception of two rows, all the non-zero entries in $\mathbf{X}$ and $\mathbf{Y}$ are on or near the digaonal: both matrices are almost banded. To solve $(\mathbf{Q} - z \mathbf{I}) \mathbf{x} = \mathbf{v}$ we multiply both sides by $\mathbf{X}$ and solve the sparse system that results. Overall, this now takes $O(K)$ time.

\subsection{Further details on the simulation studies}
\begin{table}
	\centering
	\begin{tabular}{c | c c c c c c c c l}	
		\hline		
		Simulated from & Priors & $\theta_1$  & $\theta_2$ & $\theta_3$  & $\theta_4$ & $\theta_5$ & $\theta_6$ & $\theta_{root}$ & height\\
		\hline
		$\Gamma$(2,200) & Correct  &0.91 & 0.91 & 0.94 & 0.93 & 0.94 & 0.97 & 0.95 & 1.00   \\
		$\Gamma$(2,200) & Incorrect&0.84 & 0.88 & 0.87 & 0.84 & 0.75 & 0.71 & 0.06 & 0.99\\
		\hline
		$\Gamma$(4,400) & Correct  &0.93 & 0.97 & 0.92 & 0.97 & 0.9 & 0.98 & 0.95 & 1.00 \\
		$\Gamma$(4,400) & Incorrect&0.91 & 0.95 & 0.93 & 0.94 & 0.78 & 0.68 & 0.32 & 0.99 \\
		\hline
		$\Gamma$(80,8000) & Correct  &0.98 & 0.97 & 0.99 & 0.99 & 0.97 & 0.98 & 0.97 & 1.00 \\
		$\Gamma$(80,8000) & Incorrect&0.94 & 0.95 & 0.96 & 0.93 & 0.84 & 0.83 & 0.51 & 1.00 \\
		\hline
	\end{tabular}
	\caption{Frequency that parameters fall within the $95\%$ highest posterior density of the MCMC chain for a 100 simulated datasets varying the prior on the population sizes for the trees from which the data is simulated. Note that we display the frequencies above as percentages.  }
	\label{tbl: results_sim study}
\end{table}

We conducted a simulation study to check whether the model posterior distributions recover model parameters. 
We simulate data from three different 4 species tree distributions. Parameter means are equal but we change the variance of the 4 species tree distribution. We then draw a 100 parameter sets from each 4 species tree distribution. Thereafter we simulate a 1000 SNPs for 32 individuals distributed over 4 species for each parameter set. For every simulated SNP dataset we ran two chains for 100,000 iterations, one with correct prior, i.e 4 species tree distribution used for simulation, and one with incorrect prior. See Table \ref{tbl: priors_sim_study} for details on priors. We then count the number of parameters that were recovered, i.e fall within the 95\% highest posterior density of the MCMC chain. MCMC chains took on average $\sim$200 seconds to run. We report these counts in Table \ref{tbl: results_sim study}. We see that correct priors on the model has a high rate of parameter recovery. However, when incorrect priors are used recovery rates suffer for population size parameters 
close to the root. This does not come as a surprise due to information loss on branches as we go up the tree.

\begin{table}[h]
	\begin{tabular}{clcccc}	
		\hline	
		& & \multicolumn{2}{c}{Correct prior} & \multicolumn{2}{c}{Incorrect prior}\\
		Parameter & Description & Distribution & Mean & Distribution & Mean\\
		\hline 
		$\theta$ & Population size & $\Gamma$(2, 200) & 0.01 & $\Gamma$(2, 20) & 0.1  \\
		Tree & Topology and branch length & Yule(10) &  0.0111... &  Yule(1) & 1
	\end{tabular}
\caption{We used the priors listed here to conduct the simulation study. In the case of incorrect priors, we changed the magnitude of both the Gamma and Yule priors.}
\label{tbl: priors_sim_study}
\end{table} 

\subsection{Tabulated posterior statistics for \snapp and \snapper analyses}

(Note: these could go online or in supplementary data)

\begin{center}
	\begin{longtable}{llllll}
		\caption{\snapp parameter summary after 750,000 MCMC iterations for rattlensnake dataset} \\
		\hline
		\textbf{\snapp} & \textbf{mean} & \textbf{variance} & \textbf{HPD} & \textbf{ACT} & \textbf{ESS}\\ \hline
		\endfirsthead
		\multicolumn{4}{c}%
		{\tablename\ \thetable\ -- \textit{Continued from previous page}} \\
		\hline
		\textbf{\snapp} & \textbf{mean} & \textbf{variance} & \textbf{HPD} & \textbf{ACT} & \textbf{ESS}\\
		\hline
		\endhead
		\hline \multicolumn{4}{r}{\textit{Continued on next page}} \\
		\endfoot
		\hline
		\endlastfoot 
		\hline
		theta0             & 0.1121 & 1.5601E-05 & {[}0.1042,0.1192{]}    & 1122.248  & 590.7785 \\
		theta1             & 0.0145 & 5.4621E-07 & {[}0.0129,0.0158{]}    & 1515.4263 & 437.5007 \\
		theta2             & 0.0186 & 9.4307E-07 & {[}0.017,0.0206{]}     & 3186.245  & 208.0819 \\
		theta3             & 0.023  & 1.286E-06  & {[}0.0208,0.0253{]}    & 2561.4008 & 258.8427 \\
		theta4             & 0.0103 & 5.7183E-07 & {[}8.8272E-3,0.0119{]} & 3089.9697 & 214.5652 \\
		theta5             & 0.0228 & 1.2949E-06 & {[}0.0209,0.0253{]}    & 2843.4933 & 233.1639 \\
		theta6             & 0.01   & 5.3459E-07 & {[}8.8004E-3,0.0115{]} & 2988.3364 & 221.8626 \\
		theta7             & 0.0196 & 5.7842E-07 & {[}0.0183,0.0213{]}    & 3545.1482 & 187.0162 \\
		theta8             & 0.0248 & 1.0101E-06 & {[}0.023,0.0269{]}     & 3654.5756 & 181.4164 \\
		theta9             & 0.0107 & 3.8089E-07 & {[}9.4059E-3,0.0118{]} & 3322.3719 & 199.5562 \\
		theta10            & 0.0225 & 7.0178E-07 & {[}0.0209,0.0241{]}    & 2743.8447 & 241.6317 \\
		theta11            & 0.0258 & 7.1985E-07 & {[}0.0241,0.0274{]}    & 1528.5784 & 433.7363 \\
		theta12           & 0.0669 & 1.2663E-05 & {[}0.0601,0.0735{]}    & 1035.9779 & 639.975  \\
		tree.height        & 0.0489 & 3.8258E-06 & {[}0.0452,0.0529{]}    & 1822.8776 & 363.7106
		\label{tbl:SNAPP_rattlesnake_ESS}
	\end{longtable}
\end{center}

\begin{center}
	\begin{longtable}{llllll}
		\caption{\snapper parameter summary after 750,000 MCMC iterations for rattlensnake dataset} \\
		\hline
		\textbf{\snapper} & \textbf{mean} & \textbf{variance} & \textbf{HPD} & \textbf{ACT} & \textbf{ESS}\\ \hline
		\endfirsthead
		\multicolumn{4}{c}%
		{\tablename\ \thetable\ -- \textit{Continued from previous page}} \\
		\hline
		\textbf{\snapper} & \textbf{mean} & \textbf{variance} & \textbf{HPD} & \textbf{ACT} & \textbf{ESS}\\
		\hline
		\endhead
		\hline \multicolumn{4}{r}{\textit{Continued on next page}} \\
		\endfoot
		\hline
		\endlastfoot 
		\hline
		theta0             & 0.1178    & 2.3175E-05 & {[}0.1095,0.1272{]}       & 1076.1399 & 469.2698 \\
		theta1             & 0.0117    & 4.4551E-07 & {[}0.0104,0.0131{]}       & 1833.7797 & 275.3875 \\
		theta2             & 0.0176    & 1.1315E-06 & {[}0.0156,0.0195{]}       & 1830.8958 & 275.8213 \\
		theta3             & 0.0213    & 1.4143E-06 & {[}0.019,0.0236{]}        & 1664.7158 & 303.3551 \\
		theta4             & 0.0068167 & 4.6294E-07 & {[}5.311E-3,7.9448E-3{]}  & 1783.8144 & 283.1012 \\
		theta5             & 0.0208    & 1.4231E-06 & {[}0.0185,0.023{]}        & 2257.7471 & 223.6743 \\
		theta7             & 0.0193    & 6.8242E-07 & {[}0.0179,0.0211{]}       & 2945.9165 & 171.4237 \\
		theta8             & 0.0244    & 1.1309E-06 & {[}0.0222,0.0263{]}       & 2591.8036 & 194.845  \\
		theta9             & 0.0077304 & 2.493E-07  & {[}6.8474E,3-8.7642E-3{]} & 2657.2438 & 190.0465 \\
		theta10            & 0.0228    & 9.5682E-07 & {[}0.0208,0.0247{]}       & 2374.8237 & 212.6474 \\
		theta11            & 0.028     & 1.0207E-06 & {[}0.0261,0.03{]}         & 1396.2835 & 361.6744 \\
		theta12            & 0.08      & 1.8343E-05 & {[}0.0713,0.0873{]}       & 1000      & 505      \\
		tree.height        & 0.0449    & 1.4012E-05 & {[}0.0413,0.0486{]}       & 1341.1584 & 376.5402
		\label{tbl:SNAPPER_rattlesnake_ESS}
	\end{longtable}
\end{center}

\begin{center}
\begin{longtable}{llllll}
	\caption{\snapper parameter summary after 2,000,000 MCMC iterations for freshwater turtle dataset} \\
	\hline
			\textbf{\snapper} & \textbf{mean} & \textbf{variance} & \textbf{HPD} & \textbf{ACT} & \textbf{ESS}\\ \hline
	\endfirsthead
	\multicolumn{4}{c}%
	{\tablename\ \thetable\ -- \textit{Continued from previous page}} \\
	\hline
		\textbf{\snapper} & \textbf{mean} & \textbf{variance} & \textbf{HPD} & \textbf{ACT} & \textbf{ESS}\\
	\hline
	\endhead
	\hline \multicolumn{4}{r}{\textit{Continued on next page}} \\
	\endfoot
	\hline
	\endlastfoot 
	pdist & -32288 & 58.555 & {[}-32289,-32286{]} & 6921.9588 & 237.4934 \\
	theta0 & 0.0892 & 1.09E-05 & {[}0.0829,0.0956{]} & 4559.0819 & 360.5814 \\
	theta1 & 0.0326 & 2.03E-06 & {[}0.0299,0.0354{]} & 3334.7123 & 492.972 \\
	theta2 & 0.0326 & 2.31E-06 & {[}0.0297,0.0355{]} & 4882.2185 & 336.7158 \\
	theta3 & 0.0315 & 2.28E-06 & {[}0.0288,0.0347{]} & 4594.6906 & 357.7868 \\
	theta4 & 0.089 & 1.16E-05 & {[}0.0823,0.0955{]} & 5834.7779 & 281.745 \\
	theta5 & 0.0209 & 7.38E-07 & {[}0.0195,0.0228{]} & 5032.2875 & 326.6746 \\
	theta6 & 0.0209 & 7.52E-07 & {[}0.0193,0.0227{]} & 5336.7619 & 308.037 \\
	theta7 & 0.0211 & 7.42E-07 & {[}0.0195,0.0227{]} & 5739.448 & 286.4248 \\
	theta8 & 0.0568 & 5.41E-06 & {[}0.0525,0.0613{]} & 7592.2756 & 216.5254 \\
	theta9 & 0.0576 & 5.36E-06 & {[}0.0533,0.0622{]} & 5374.5587 & 305.8706 \\
	theta10 & 0.0576 & 5.40E-06 & {[}0.0535,0.0621{]} & 5700.7353 & 288.3698 \\
	theta11 & 0.0398 & 3.51E-06 & {[}0.0362,0.0437{]} & 4824.2314 & 340.763 \\
	theta12 & 0.0475 & 2.82E-06 & {[}0.0439,0.0506{]} & 5591.2287 & 294.0176 \\
	theta13 & 0.0474 & 3.08E-06 & {[}0.0441,0.0508{]} & 7868.6887 & 208.9192 \\
	theta14 & 0.0473 & 3.13E-06 & {[}0.0441,0.0509{]} & 6942.6134 & 236.787 \\
	theta15 & 0.0474 & 2.99E-06 & {[}0.0441,0.0507{]} & 7085.1136 & 232.0246 \\
	theta16 & 0.0473 & 2.84E-06 & {[}0.0442,0.0503{]} & 4927.4417 & 333.6254 \\
	theta17 & 0.0563 & 5.91E-06 & {[}0.0512,0.0608{]} & 6188.3147 & 265.649 \\
	theta18 & 0.0254 & 1.90E-06 & {[}0.0229,0.0283{]} & 4461.407 & 368.4756 \\
	theta19 & 0.021 & 1.58E-06 & {[}0.0186,0.0233{]} & 3920.8397 & 419.2776 \\
	theta20 & 0.0351 & 2.47E-06 & {[}0.032,0.0381{]} & 5149.6604 & 319.2288 \\
	theta21 & 0.035 & 2.45E-06 & {[}0.032,0.0382{]} & 5103.4682 & 322.1182 \\
	theta22 & 0.0312 & 1.98E-06 & {[}0.0286, 0.034{]} & 5843.4713 & 281.326 \\
	theta23 & 0.0275 & 1.68E-06 & {[}0.0249,0.0298{]} & 4633.1697 & 354.8154 \\
	theta24 & 0.0137 & 5.95E-07 & {[}0.0121,0.0151{]} & 6877.4641 & 239.03 \\
	theta25 & 0.0137 & 6.94E-07 & {[}0.0121,0.0152{]} & 7605.5597 & 216.1472 \\
	theta26 & 0.0105 & 3.66E-07 & {[}9.3707E-3,0.0116{]} & 5984.3981 & 274.701 \\
	theta27 & 0.035 & 1.34E-06 & {[}0.0328,0.0371{]} & 4350.192 & 377.896 \\
	theta28 & 0.0324 & 1.52E-06 & {[}0.0301,0.0349{]} & 6937.3698 & 236.9658 \\
	theta29 & 0.0324 & 1.51E-06 & {[}0.03,0.0347{]} & 6434.1179 & 255.5004 \\
	theta30 & 0.033 & 1.42E-06 & {[}0.0305,0.0352{]} & 5917.0715 & 277.8266 \\
	theta31 & 0.033 & 1.34E-06 & {[}0.0304,0.035{]} & 5047.0492 & 325.719 \\
	theta32 & 0.0346 & 1.40E-06 & {[}0.0325,0.0371{]} & 5272.0715 & 311.8168 \\
	theta33 & 0.0345 & 1.55E-06 & {[}0.0321,0.0367{]} & 4934.2925 & 333.1622 \\
	theta34 & 0.0346 & 1.52E-06 & {[}0.0322,0.037{]} & 5797.7463 & 283.5446 \\
	theta35 & 0.0345 & 1.39E-06 & {[}0.0319,0.0366{]} & 5713.2156 & 287.7398 \\
	theta36 & 0.0343 & 1.49E-06 & {[}0.0321,0.0367{]} & 5738.9424 & 286.45 \\
	theta37 & 0.032 & 1.73E-06 & {[}0.0296,0.0346{]} & 5566.4597 & 295.326 \\
	theta38 & 8.24E-03 & 1.83E-07 & {[}7.3532E-3,9.0356E-3{]} & 4660.0198 & 352.771 \\
	theta39 & 0.0285 & 2.06E-06 & {[}0.0254,0.031{]} & 2754.3786 & 596.8388 \\
	theta40 & 0.045 & 3.94E-06 & {[}0.0416,0.0493{]} & 2735.1968 & 601.0244 \\
	theta41 & 0.0891 & 1.03E-05 & {[}0.0823,0.0946{]} & 4363.7001 & 376.7262 \\
	theta42 & 0.0328 & 1.74E-06 & {[}0.0302,0.0352{]} & 4046.0439 & 406.303 \\
	theta43 & 0.0344 & 1.63E-06 & {[}0.0321,0.037{]} & 4026.7212 & 408.2528 \\
	theta44 & 0.0553 & 2.70E-06 & {[}0.0522,0.0585{]} & 7152.4991 & 229.8386 \\
	theta45 & 0.0209 & 7.09E-07 & {[}0.0196,0.023{]} & 4865.2479 & 337.8902 \\
	theta46 & 0.021 & 6.24E-07 & {[}0.0195,0.0226{]} & 4733.2039 & 347.3166 \\
	theta47 & 0.0518 & 4.84E-06 & {[}0.0481,0.0555{]} & 26723.5617 & 61.5158 \\
	theta48 & 0.0575 & 4.88E-06 & {[}0.0538,0.0621{]} & 5530.8802 & 297.2258 \\
	theta49 & 0.0573 & 4.28E-06 & {[}0.0535,0.0612{]} & 6291.947 & 261.2736 \\
	theta50 & 0.0567 & 3.64E-06 & {[}0.0531,0.0607{]} & 7435.9237 & 221.0782 \\
	theta51 & 0.0475 & 2.81E-06 & {[}0.0441,0.0506{]} & 7190.8831 & 228.6116 \\
	theta52 & 0.0474 & 2.92E-06 & {[}0.0442,0.0506{]} & 6880.0099 & 238.9416 \\
	theta53 & 0.0473 & 2.65E-06 & {[}0.0444,0.0507{]} & 5266.9057 & 312.1226 \\
	theta54 & 0.0473 & 2.64E-06 & {[}0.044,0.0502{]} & 5930.9334 & 277.1772 \\
	theta55 & 0.0452 & 2.35E-06 & {[}0.0423,0.0481{]} & 6519.7251 & 252.1456 \\
	theta56 & 0.0531 & 2.40E-06 & {[}0.0502,0.0562{]} & 10545.795 & 155.884 \\
	theta57 & 0.0246 & 1.12E-06 & {[}0.0225,0.0265{]} & 6463.5958 & 254.3352 \\
	theta58 & 0.0351 & 2.26E-06 & {[}0.0324,0.0382{]} & 4395.8089 & 373.9744 \\
	theta59 & 0.0348 & 1.46E-06 & {[}0.0326,0.0372{]} & 12531.3457 & 131.1846 \\
	theta60 & 0.0387 & 1.12E-06 & {[}0.0366,0.0405{]} & 16803.2068 & 97.8338 \\
	theta61 & 0.0276 & 1.16E-06 & {[}0.0254,0.0294{]} & 11382.032 & 144.4312 \\
	theta62 & 0.0136 & 5.75E-07 & {[}0.0122,0.015{]} & 7378.0828 & 222.8112 \\
	theta63 & 0.0132 & 3.93E-07 & {[}0.012,0.0144{]} & 9066.0779 & 181.3264 \\
	theta64 & 0.025 & 7.92E-07 & {[}0.0232,0.0266{]} & 13972.7813 & 117.6516 \\
	theta65 & 0.0492 & 2.12E-06 & {[}0.0465,0.052{]} & 24561.4047 & 66.931 \\
	theta66 & 0.0345 & 1.30E-06 & {[}0.0324,0.0367{]} & 4752.1445 & 345.9322 \\
	theta67 & 0.0344 & 1.22E-06 & {[}0.0323,0.0366{]} & 4528.4154 & 363.0232 \\
	theta68 & 0.0345 & 1.28E-06 & {[}0.0324,0.0368{]} & 5549.8598 & 296.2092 \\
	theta69 & 0.0344 & 1.21E-06 & {[}0.0323,0.0364{]} & 4893.2052 & 335.9598 \\
	theta70 & 0.0343 & 1.13E-06 & {[}0.0323,0.0363{]} & 4619.4067 & 355.8726 \\
	theta71 & 0.0324 & 1.44E-06 & {[}0.0301,0.0346{]} & 6038.6503 & 272.233 \\
	theta72 & 0.033 & 1.33E-06 & {[}0.0304,0.035{]} & 5481.3538 & 299.9114 \\
	theta73 & 0.0329 & 1.25E-06 & {[}0.0305,0.0349{]} & 5176.6725 & 117.563 \\
	theta74 & 0.0336 & 1.09E-06 & {[}0.0317,0.0357{]} & 4940.5417 & 332.7408 \\
	theta75 & 0.0334 & 1.06E-06 & {[}0.0314,0.0354{]} & 5603.2564 & 293.3866 \\
	theta76 & 0.037 & 3.01E-06 & {[}0.0336,0.0401{]} & 22711.2203 & 72.3836 \\
	theta77 & 0.0492 & 4.41E-06 & {[}0.0456,0.0533{]} & 33200.5358 & 49.5148 \\
	theta78 & 0.06 & 5.72E-06 & {[}0.0558,0.064{]} & 34594.3404 & 47.52 \\
	theta79 & 0.0383 & 9.33E-06 & {[}0.0328,0.0438{]} & 25502.3797 & 64.4614 \\
	theta80 & 0.0531 & 3.43E-06 & {[}0.0495,0.0565{]} & 33045.9805 & 49.7464 \\
	tree.height & 0.1982 & 3.96E-05 & {[}0.1863,0.2107{]} & 1811.7552 & 907.3632 \\
	tree.length & 0.754 & 2.41E-04 & {[}0.7242,0.7845{]} & 2135.0573 & 769.9652 \\
	\label{tbl: SNAPPER_turtles_ESS}
\end{longtable}
\end{center}

\end{document}